\documentclass[grl]{agutex}

\usepackage[dvips,dvipdf]{graphicx}

\authorrunninghead{SUBRAMANIAN, LARA \& BORGAZZI}

\titlerunninghead{SW-ICME VISCOUS DRAG}

\authoraddr{Prasad Subramanian, Indian Institute of Science Education and Research, Sai Trinity Building, Garware Circle, Pashan, Pune - 411021, India (p.subramanian@iiserpune.ac.in)}

\authoraddr{Alejandro Lara and Andrea Borgazzi, Instituto de Geof\'isica, Universidad Nacional Aut\'onoma de M\'exico, C. U. M\'exico D.F 04510, M\'exico}

\begin{document}


%


%


\title{Can solar wind viscous drag account for CME deceleration?}

%



%


%


%





\authors{Prasad Subramanian,\altaffilmark{1}
Alejandro Lara,\altaffilmark{2} and Andrea Borgazzi\altaffilmark{2}}

\altaffiltext{1}{Indian Institute of Science Education and Research, Pune, India}

\altaffiltext{2}{Instituto de Geof\'isica, Universidad Nacional Aut\'onoma de M\'exico, M\'exico}








%


%




\begin{abstract}
The forces acting on solar Coronal Mass Ejections (CMEs) in the interplanetary medium have been evaluated so far in terms of an empirical drag coefficient $C_{\rm D} \sim 1$ that quantifies the role of the aerodynamic drag experienced by a typical CME due to its interaction with the ambient solar wind. We use a microphysical prescription for viscosity in the turbulent solar wind to obtain an analytical model for the drag coefficient $C_{\rm D}$. This is the first physical characterization of the aerodynamic drag experienced by CMEs. We use this physically motivated prescription for $C_{\rm D}$ in a simple, 1D model for CME propagation to obtain velocity profiles and travel times that agree well with observations of deceleration experienced by fast CMEs. 

\end{abstract}


%


%



%



\begin{article}


%


%


\section{Background}
The aerodynamic drag experienced by CMEs as they traverse the interplanetary medium between the Sun and Earth is generally thought to arise due to the coupling of the CMEs to the ambient solar wind. CMEs which start out slow (with respect to the solar wind speed) near the Sun seem to accelerate en route to the Earth, while fast CMEs are decelerated, suggesting that the solar wind strongly mediates CME propagation (Gopalswamy et al 2000, Manoharan 2006) in the interplanetary medium. This fact has been invoked in several papers that derive a heuristic aerodynamic drag coefficient for CMEs (e.g., Byrne et al 2010; Maloney \& Gallagher 2010; Vr\u{s}nak et al 2010; 2012; Cargill 2004). In particular, Borgazzi et al (2009) have used two different drag prescriptions to investigate CME slowdown using a simple 1D hydrodynamical model that lends itself to analytical solutions.

While there has been a fair amount of progress in this direction, a physical understanding of the viscosity mechanism that leads to the drag on CMEs is still lacking. Many authors have focussed on the overall dynamics of the CME, using the drag coefficient $C_{D}$ only as an empirical fitting parameter that remains constant throughout the propagation (e.g., Byrne et al 2010; Maloney \& Gallagher 2010; Lara et al 2011). 

\section{This work}
Our primary focus here is therefore on computing the drag force on the CME using a physical prescription. We compute the viscosity of the ambient solar wind using a prescription relevant to collisionless plasmas. 
We use this solar wind viscosity prescription to compute the drag on the expanding CME employing approaches that are commonly used in fluid dynamics. We restrict ourselves to a simple 1D, hydrodynamical model in order to focus on the essential physics. We first recapitulate the basics of the drag force experienced by CMEs.

\section{Viscous drag on a CME}
We start with the oft-used 1D equation of motion for a CME that experiences only aerodynamic drag (e.g., Borgazzi et al 2009):
\begin{equation}
m_{\rm CME}\,V_{\rm CME}\,\frac{d V_{\rm CME}}{d R} = \frac{1}{2}\,C_{\rm D} \,N_{i}\,m_{p}\,A_{\rm CME}\,(V_{\rm CME} - V_{\rm sw})^{2}\, ,
\label{eq14}
\end{equation}
where $m_{\rm CME}$ is the CME mass, $V_{\rm CME}$ is the CME speed, $C_{\rm D}$ is the all-important dimensionless drag coefficient, $N_{i}$ is the proton number density in the ambient solar wind, $m_{p}$ is the proton mass, $A_{\rm CME} \equiv \pi R_{\rm CME}^{2}$ is the cross-sectional area of the CME and $V_{\rm sw}$ is the solar wind speed. The term on the right hand side of Eq~(\ref{eq14}) represents the drag force experienced by the CME. It may be noted that this form for the drag force is appropriate only for a {\em solid} body moving through a medium at high Reynolds numbers (e.g., Landau \& Lifshitz 1987).   
Most authors adopt an empirical drag coefficient $C_{\rm D}$ that remains constant with heliocentric distance. The value for $C_{\rm D}$ is chosen to ensure that the computed velocity profile using an equation like Eq~\ref{eq14} agrees with observations (e.g., Lara et al 2011).

It is well known, however, that the drag coefficient $C_{\rm D}$ is a function of the Reynolds number of the system under consideration. We start with data from Achenbach (1972) who gives the widely used characterization of $C_{\rm D}$ as a function of the Reynolds number for high Reynolds number flow past a solid sphere. This standard characterization can be found in most fluid dynamics texts (e.g., Landau \& Lifshitz 1987). The drag coefficient $C_{\rm D}(Re)$ is a very slowly increasing function of the Reynolds number $Re$ until $Re \approx 10^5$, above which it exhibits a sharp drop; this is the so-called drag crisis. Beyond this sharp drop, $C_{\rm D}$ is an increasing function of $Re$; this spans the supercritical and the transcritical regimes. While these facts are well known, we are not aware of any analytical fits to Achenbach's (1972) data, especially for high Reynolds numbers. We therefore fit Achenbach's (1972) data for $Re > 10^6$ to the following functional form:
\begin{equation}
C_{\rm D}(Re) = 0.1478 - \frac{42834}{Re} + 9.8 \times 10^{-9} Re \, .
\label{cdform}
\end{equation}

In our case, the Reynolds number is a function of the CME velocity $V_{\rm CME}$, the typical macroscopic lengthscale $R_{\rm CME}$ and the viscosity coefficient according to the usual formula 
\begin{equation}
Re \equiv \frac{V_{\rm CME}\,R_{\rm CME}}{\nu} = \frac{V_{\rm CME}\,R_{\rm CME}\,N_{i}\,m_{p}}{\eta}\, ,
\label{Rey}
\end{equation}
where $\nu$ (${\rm cm^{2}\,s^{-1}}$) is the coefficient of kinematic viscosity and $\eta \equiv \nu\,N_{i}\,m_{p}$.
Our use of the CME radius $R_{\rm CME}$ for a typical macroscopic lengthscale implies that (as for a flux rope CME) the cross-sectional dimensions of the CME are $\approx R_{\rm CME}$.
In what follows, we will develop a physically motivated viscosity prescription for the solar wind and use it to determine the Reynolds number self-consistently. 
We will use the Reynolds number thus determined to calculate $C_{\rm D}$ and use it to solve the CME equation of motion (Eq~\ref{eq14}).

\section{Viscosity prescription for a collisionless plasma}
It is well known that the solar wind is a collisionless plasma above $\sim$ 10 $R_{\odot}$, where the (Coulomb) collision mean free path is longer than macroscopic scale lengths. However, a fluid description is a generally well accepted one for the solar wind. This implies that the collisionless particles are confined via effective collisions with scattering centers of some sort, which lends validity to a fluid treatment. Clearly, viscosity in the solar wind is not due to interparticle collisions, as is the case in everyday experience. 

\subsection{Hybrid viscosity}
We follow Subramanian, Becker \& Kafatos (1996; SBK96 from now on), who developed a model for collisionless viscosity. They considered a one-dimensional, plane-parallel shear flow, with viscosity being provided by the flux of protons that originate from one of the layers and impinge on the other. Tangled magnetic fields are embedded in the flow, and the average coherence length (i.e., the length for which an average magnetic field line is expected to remain straight) of the ``kinks'' in the magnetic fields is taken to be $\lambda_{\rm coh}$.
The protons, whose gyroradii are negligible in comparison to other length scales, are envisaged to slide along the field lines, and change their direction (i.e., get scattered) either when they encounter another proton, or a kink in the tangled magnetic field. Since the protons are collisionless, proton-proton collisions are unlikely, and momentum transfer (and consequently viscosity) is dominated by proton-magnetic kink encounters. The magnetic kinks can also be envisaged as turbulent eddies that act as scattering centers; the only restriction in this scenario is that the scattering centers evolve over a timescale that is slower than that of the travel time of a typical proton, so that they appear stationary to it. Using this scenario, SBK96 arrive at the following simple and physically motivated formula for a ``hybrid'' coefficient of dynamic viscosity $\eta_{{\rm hyb}}$; one that is neither due to magnetic field stresses alone, nor due to proton-proton (Coulomb) collisions alone:

\begin{equation}
\eta_{{\rm hyb}} \equiv \nu_{\rm hyb}\,N_{i}\,m_{p} = \frac{2}{15} \frac{\lambda}{\lambda_{\rm ii}} \, \eta_{\rm ff} \,\,\,\,\,\,\,\,\, {\rm g\, cm^{-1}\, s^{-1}} \, ,
\label{eq1}
\end{equation}
where $\lambda$ denotes the effective mean free path, $\lambda_{\rm ii}$ is the (Coulomb) mean free path for proton-proton collisions and $\eta_{\rm ff}$ is the standard proton viscosity due to Coulomb collisions alone (Spitzer 1962). The viscosity $\eta_{\rm hyb}$ in a collisionless plasma is thus suppressed with respect to the standard Coulomb value $\eta_{\rm ff}$ by a factor $(2/15) \lambda/\lambda_{\rm ii}$. 
For the sake of completness, we reproduce the expressions for $\lambda_{\rm ii}$ and $\eta_{\rm ff}$:
{\begin{eqnarray}
\nonumber\lambda_{\rm ii} = v_{\rm rms} t_{\rm ii} = 1.8 \times 10^{5} \frac{T_{i}^{2}}{N_{i} \, {\rm ln} \Lambda}\,\,\,\,\,\,\,\,\,{\rm cm} \\
\eta_{\rm ff} = \sqrt{6}\,N_{i}\,m_{p}\,v_{\rm rms}\,\lambda_{\rm ii} =  2.2 \times 10^{-15} \, \frac{T_{i}^{5/2}}{{\rm ln} \Lambda}\,\,\,\,\,\,\,\,\, {\rm g\, cm^{-1}\, s^{-1}} \, ,
\label{eq2}
\end{eqnarray}}
where $v_{\rm rms} \equiv (3 k T_{i}/m_{p})^{1/2}$ is the rms thermal velocity of the protons, $t_{\rm ii}$ is the mean interval between proton-proton (Coulomb) collisions, $N_{i}$ is the proton number density in ${\rm cm}^{-3}$, $T_{i}$ is the proton temperature in Kelvin and ${\rm ln} \Lambda$ is the Coulomb logarithm, which is taken to be equal to 20. 
Eqs~(\ref{eq1}) and (\ref{eq2}) comprise our operational definition for hybrid viscosity, which we will apply to the solar wind. It may be noted that the hybrid viscosity (Eq~\ref{eq1}) can be equivalently written as
\begin{equation}
\eta_{\rm hyb} = \frac{2}{15}\,\sqrt{6}\,N_{i}\,m_{p}\,v_{\rm rms}\,\lambda\, .
\label{eq1b}
\end{equation}
Eq~\ref{eq1b} looks similar to the usual fluid dynamical expression for turbulent viscosity (e.g., Landau \& Lifshitz 1987)
\begin{equation}
\eta_{\rm turb} \approx N_{i}\,m_{p}\, v_{\rm turb}\,l_{\rm turb} \, ,
\label{eq1b1}
\end{equation}
where $v_{\rm turb}$ is the velocity of the turbulent eddies of lengthscale $l_{\rm turb}$. Furthermore, if $l_{\rm turb}$ is interpreted as the dissipation lengthscale in a turbulent cascade, and if we adopt the usual expression for energy dissipation rate in Kolmogorov turbulence $\epsilon \sim v_{\rm turb}^{3}/l_{\rm turb}$, the expression for turbulent viscosity can be rewritten as (Verma 1996)
\begin{equation}
\eta_{\rm turb} \approx N_{i}\,m_{p}\,(\epsilon\,l_{\rm turb}^{4})^{1/3} \, .
\label{eq1c}
\end{equation}

\subsection{Solar wind viscosity}
We now use Eqs~(\ref{eq1}) and (\ref{eq2}) to compute the operative viscosity in the ambient solar wind. 
As mentioned earlier, the viscosity would be determined primarily by collisions between protons and magnetic kinks, and therefore the effective mean free path $\lambda \approx \lambda_{\rm coh}$. The coherence length of the magnetic field irregularities ($\lambda_{\rm coh}$) represents the shortest lengthscale over which the turbulent magnetic field is structured.
It is fairly well established that density turbulence in the solar wind follows the Kolmogorov scaling, with an inner (dissipation) scale that is determined by proton cyclotron resonance (Coles \& Harmon 1989). We assume that the magnetic field irregularities follow the density irregularities (e.g., Spangler 2002), and are governed by the same inner scale. We therefore take 
$\lambda_{\rm coh}$ to be equal to the inner scale of solar wind density turbulence advocated by Coles \& Harmon (1989):

\begin{equation}
\lambda \approx \lambda_{\rm coh} = 684\, N_{i}^{-1/2}\,\,\,\,\,\,\,\,\,{\rm km}\, .
\label{eq3}
\end{equation}

The solar wind proton density $N_{i}$ is assumed to be given by the model of LeBlanc et al (1998): \begin{equation} \label{eq:leblanc}
N_{i}(R) = 3.3 \times 10^{5}\,R^{-2} + 4.1 \times 10^{6}\,R^{-4} + 8 \times 10^{7}\,R^{-6}\,\,\,\,\,{\rm cm^{-3}}\, ,
\end{equation}
where $R$ is the heliocentric distance in solar radii. 
We take the proton temperature to be $T_i = 10^5$ K.

Thus Eq~(\ref{eq1}), with $\lambda$ given by Eq~(\ref{eq3}) and the density given by Eq~(\ref{eq:leblanc}) defines the hybrid viscosity prescription for the ambient solar wind for our purposes. This enables us to obtain the viscosity of the ambient solar wind as a function of heliocentric distance.

Eviatar \& Wolf (1968) obtain an estimate of $\nu \approx 800 \, {\rm km}^{2}\,{\rm s}^{-1}$ at the Earth by considering the momentum transfer across the magnetopause due to fluctuations induced by the two-stream cyclotron instability. This is in excellent agreement with the value of 788 ${\rm km}^{2}\,{\rm s}^{-1}$ we obtain for the kinematic viscosity coefficient from the hybrid viscosity model. Perez-de-Tejada (2005) has derived a rough estimate of $\nu  \approx 1000 \, {\rm km}^{2}\,{\rm s}^{-1}$ for the kinematic viscosity of the solar wind near the ionosheath of Venus (heliocentric distance 0.72 AU). The hybrid viscosity model yields a value of 600 ${\rm km}^{2}\,{\rm s}^{-1}$ for the coefficient of dynamic viscosity at 0.72 AU.
\section{Results}
We use the prescription for the viscosity given in \S~4 to determine the Reynolds number (Eq~\ref{Rey}), which in turn is used to determine $C_{\rm D}$ (Eq~\ref{cdform}). The resulting expression for $C_{\rm D}$ is used in the equation of motion (Eq~\ref{eq14}) to solve for the CME speed as a function of heliocentric distance.

A representative result is shown in Figures 1 and 2. In this example, the CME radius $R_{\rm CME}$ is assumed to expand as $R_{\rm CME} = K R^{p}$, where $K$ is a constant of proportionality, which we determine by assuming that the radius of the CME is 1 $R_{\odot}$ at a heliocentric distance of 2 $R_{\odot}$. The power law index $p$ is assumed to be equal to 0.78 (Bothmer \& Schwenn 1998). The solar wind speed $V_{\rm sw}$ is taken to be 375 km/s and the velocity of the CME at $R = 5 R_{\odot}$ is taken to be equal to 1048 km/s. This CME initial velocity is representative of a CME on May 17 2008 that was well observed by the STEREO spacecraft from the Sun to the Earth (Wood et al 2009). We chose this CME since it is one of the few 
fast ones that were well observed from the Sun to the Earth by the STEREO spacecraft during the minimum and ascending phases of cycle 24. 
The CME mass $m_{\rm CME}$ is taken to be $5 \times 10^{14}$ g. This value for the CME mass is representative of CMEs during this part of the solar cycle (see Vourlidas et al 2010).

The solid line in Fig 1 is the solution for the velocity profile calculated from Eq~(\ref{eq14}). The plus signs represent data for the CME of May 17 2008. In order to get the velocity-distance data for this CME, we started with the distance-time data for this CME (top panel, Fig 6 of Wood et al 2009). The velocity-distance data points in Fig 1 are derived by numerically differentiating the distance-time data. The large undulations in the velocity-distance data shown in Fig 1, especially for distances $> 20 R_{\odot}$, are 
due to fluctuations in the distance-time measurements,
which are accentuated in the derived velocity. In particular, the significant dip in CME speed after $\sim$ 150 $R_{\odot}$ is almost certainly an unphysical artefact. It is well known that CMEs achieve an asymototic speed by the time they reach the Earth, and often as soon as $\sim$ 100 $R_{\odot}$ (e.g., Poomvises, Zhang \& Olmedo 2010). The dip in the CME speed beyond 150 $R_{\odot}$ is probably due to the difficulty in tracking CME in HI data, which are well known to have poor signal to noise ratio at those distances. In view of this, the agreement between the solid line, which is the solution to Eq~(\ref{eq14}) and the velocity-distance data for the May 17 2008 CME is evidently rather good. The Reynolds number for the CME (in units of $10^{8}$) and the dimensionless constant $C_{\rm D}$ (which is now computed self-consistently, using the prescription for solar wind viscosity) are shown in Fig 2.
The results shown in Fig 2 justify the function we adopt for $C_{\rm D}$ (Eq~\ref{cdform}), which is valid only for Reynolds number $\gg 10^{4}$. We note that Wood et al (2009) obtain the velocity for the May 17 2008 not by piecewise numerical differentiation of the distance-time data (as we have done), but by assuming an empirical model that incorporates an initial acceleration phase and a subsequent deceleration phase that is followed by a constant velocity phase.

We have also carried out this exercise for some of the fastest Earth-directed halo CMEs observed so far. Our results are summarized in Table 1. For each event, the parameters supplied to the model (Eq~\ref{eq14}) are the solar wind speed $V_{\rm SW}$, the CME mass $m_{\rm CME}$ and the CME initial velocity $V_{i}$. The initial velocity of each halo CME is determined from the LASCO CME catalog (http://cdaw.gsfc.nasa.gov/CME$_{-}$list/). Since these are halo CMEs, their masses are hard to estimate, and we have used reasonable guesses, based on typical CME masses during the appropriate phase in the solar cycle (Vourlidas et al 2010). 
As before, the proton temperature is assumed to be $10^{5}$ K and the Leblanc et al (1998) density model (Eq~\ref{eq:leblanc}) is used for determining the solar wind viscosity and the drag coefficient $C_{\rm D}$. The quantity $V_{\rm ICME}$ represents the speed of the relevant interplanetary CME detected in-situ by spacecraft near the Earth and $TT_{\rm ICME}$ is the time elapsed between the first detection of the halo CME in the LASCO FOV and the detection of the corresponding ICME at the Earth. The quantity $V_{\rm model}$ represents the predicted ICME velocity at the Earth and $TT_{\rm model}$ denotes the predicted Sun-Earth travel time. 
It is evident from Table 1 that the model predictions for the ICME speed at the Earth and the total travel time agree quite well with the observations.  
\begin{table*}
\caption{A comparison between observations and model results for near-Earth ICME speeds and Sun-Earth travel times for some fast Earth-directed CMEs.}\label{TA11}
\centering
\begin{tabular}{ccccccccc}
\hline \hline
Event & $V_{\rm SW}$  & $m_{\rm CME}$ & $V_{i}$ & $V_{\rm ICME}$ & $V_{\rm model}$ & $TT_{\rm ICME}$ & $TT_{\rm model}$  \\
\hline
 & (km/s) & (g) & (km/s) & (km/s) & (km/s) & hours & hours \\
\hline
\hline
20010409 & 450 & $5 \times 10^{14}$ & 1192 & 670 & 696 & 53.4 & 53.5 \\
20031028 & 450 &$2 \times 10^{15}$ & 2459 & 1400 & 1350 & 24.8 & 25.6 \\
20040726 & 400 & $10^{15}$ & 1366 & 900 & 894 & 38.63 & 37 \\
\hline \hline
\end{tabular}
\end{table*} 

We have used the first CME in Table 1 (20010409) as a representative event to ascertain the effects of varying some of the parameters on the model predictions. All other quantities remaing fixed, we find that a 10 \% decrease (increase) in $V_{\rm sw}$ results in a 7.5 \% decrease (increase) in $V_{\rm model}$ and a 7 \% increase (decrease) in $TT_{\rm model}$. Similarly, we find that a 50 \% decrease (increase) in $m_{\rm CME}$ leads to a 13 \% decrease (increase) in $V_{\rm model}$ and a 15 \% increase (decrease) in $TT_{\rm model}$. On the other hand, a 100 \% increase (decrease) in $T_{i}$ results in a 4 \% increase (decrease) in $V_{\rm model}$ and a 4.6 \% decrease (increase) in $TT_{\rm model}$.
Therefore, the results of our model are most (least) sensitive to the solar wind speed (proton temperature).

\begin{figure}
\includegraphics[width=0.5 \textwidth]{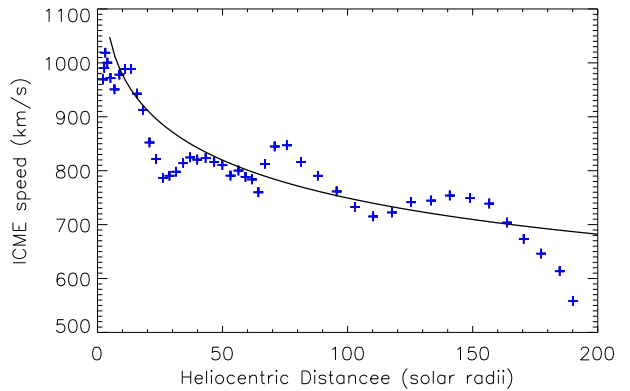}
\caption{The solid line denotes the model prediction for the CME velocity (in km/s) as a function of heliocentric distance (in $R_{\odot}$). The plus signs are derived from the distance-time data for the May 17 2008 CME. Further details are mentioned in the text.}
\end{figure}

\begin{figure}
\includegraphics[width=0.5 \textwidth]{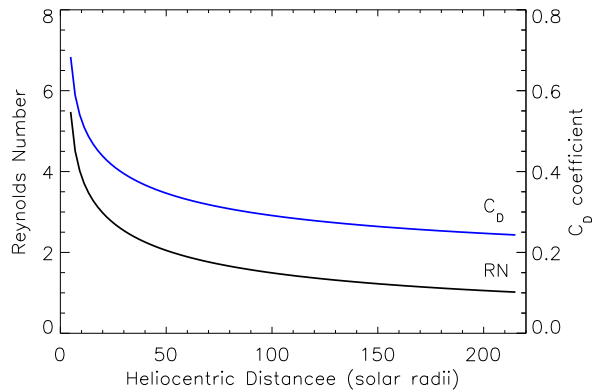}
\caption{Heliocentric variation of the CME Reynolds number in multiples of $10^8$ ($RN$) and the drag coefficient ($C_{\rm D}$) for the model of Fig 1.}
\end{figure}

\section{Summary and Conclusions}
\subsection{Summary}
To summarize, we have considered the motion of a CME under the influence of {\em only} a drag force (Eq~\ref{eq14}). The drag force is proportional to the square of the CME speed (relative to the background solar wind); this form is appropriate for a solid body moving through a background medium at high Reynolds numbers. The proportionality constant involves the important (dimensionless) drag coefficient $C_{\rm D}$, which has been treated as an empirical fitting parameter in the literature so far.

It is, however, well known that the drag coefficient $C_{\rm D}$ is in fact a function of the CME Reynolds number $Re$. We use standard data pertaining to the motion of a solid sphere at high Reynolds numbers to characterize $C_{\rm D}$ as a function of $Re$ (Eq~\ref{cdform}). It is necessary to know the coefficient of kinematic viscosity $\nu$ in order to determine $Re$ (Eq~\ref{Rey}). We compute the operative viscosity in the (quiescent) collisionless solar wind using a prescription that considers protons colliding with magnetic scattering centers (Eqs~\ref{eq1} and \ref{eq2}). We assume that the proton temperature in the solar wind is $T_{i} = 10^5$K, the proton number density is given by Eq~(\ref{eq:leblanc}) and that the operative mean free path $\lambda$ is the inner scale of the spectrum of density irregularities (Eq~\ref{eq3}). This gives a simple, physically motivated prescription for the drag coefficient $C_{\rm D}$, which we use in solving a simplified 1D equation of motion (Eq~\ref{eq14}). Using observational estimates for the CME starting speed and reasonable guesses for the ambient solar wind speed and the CME mass, we find that our results for the CME speed profile and the Sun-Earth travel time agree quite well with observations (Fig 1 and Table 1).

\subsection{Conclusions}
The agreement between our theoretical predictions and observations of CME deceleration is remarkable, especially in light of the several simplifying assumptions we have adopted. It confirms the essential validity of our physical prescription for the dimensionless viscous drag parameter $C_{\rm D}$, which can be profitably used in semi-analytical treatments (e.g., Vr\u{s}nak et al 2012) and simulations of CME propagation.

We next mention some caveats that accompany our conclusions.
Firstly, this is a rather simplified, 1D hydrodynamic treatment that only addresses the essential physics of the CME-solar wind interaction. There is no attempt to include Lorentz force driving of CMEs; something that is known to be important upto around 30 $R_{\odot}$ (e.g., Subramanian \& Vourlidas 2007), at least for CMEs that are only moderately fast. Very fast CMEs, such as the ones considered in Table 1, probably experience Lorentz force driving and acceleration very early in their evolution. Upcoming instruments such as the ADITYA-I coronograph (Singh et al 2011) could address this issue via high cadence images of the inner solar corona.

Secondly, it is surprising that a drag law of the form of Eq~(\ref{eq14}), which is in fact valid only for solid bodies, works so well. CMEs are often thought of as {\em bubbles}, which are technically defined as bodies that deform in response to tangential stresses on their surfaces. The drag force for high Reynolds number flow past a bubble is in fact proportional to $V_{\rm CME}$, and not to $V_{\rm CME}^{2}$ (Landau \& Lifshitz 1987; Merle et al 2005). With this form for the drag force, our preliminary results indicate that a typical fast CME slows down only by about 1 \% by the time it reaches the Earth. It is therefore worth investigating why CMEs seem to behave like solid bodies as far as their interaction with the ambient solar wind is concerned. In a flux rope CME, the large-scale, ordered magnetic field of the flux rope could provide an explanation for this apparent ``solid body'' effect. Finally, we note that the drag coefficient derived from ideal MHD simulations (e.g., Cargill 2004) can often significantly exceed unity, in contrast to the values obtained in this work.






%




%




%


%


%






%








%


\begin{acknowledgments}
We acknowledge several comments and suggestions from the anonymous referees that have helped improve this paper. PS acknowledges financial support from the RESPOND program of the Indian Space Research Agency.
AL acknowledges partial support from DGAPA-UNAM IN112412-3 and CONACyT grants.

\end{acknowledgments}

\end{article}




%


%





%






%




%



%





%




%



%





















\end{document}